\documentclass{elsart4-1}

\usepackage{graphicx}

\usepackage{amssymb}

\usepackage[english,francais]{babel}

\newtheorem{e-proposition}[theorem]{Proposition}

\newtheorem{e-definition}[theorem]{Definition\rm}

\renewcommand{\theequation}{\arabic{equation}}
\def\og{\leavevmode\raise.3ex\hbox{$\scriptscriptstyle\langle\!\langle$~}}
\def\fg{\leavevmode\raise.3ex\hbox{~$\!\scriptscriptstyle\,\rangle\!\rangle$}}

\newcommand{\jb}{\mbox{\boldmath $j$}}

\newcommand{\be}{\begin{equation}}
\newcommand{\ee}{\end{equation}}
\newcommand{\bea}{\begin{eqnarray}}
\newcommand{\eea}{\end{eqnarray}}
\newcommand{\les}{\ell_{ES}}
\newcommand{\um}{\mbox{${1\over 2}$}}

\begin{document}

\centerline{Physics or Astrophysics/Header}
\begin{frontmatter}


\selectlanguage{english}
\title{Surface currents and slope selection in crystal growth}


\selectlanguage{english}
\author{Paolo Politi}
\ead{Paolo.Politi@isc.cnr.it}
\ead[url]{http://www.ifac.cnr.it/\~{}politi}

\address{Istituto dei Sistemi Complessi,
Consiglio Nazionale delle Ricerche, Via Madonna del Piano 10,
50019 Sesto Fiorentino, Italy}


\medskip
\begin{center}
{\small Received *****; accepted after revision +++++}
\end{center}

\begin{abstract}
We face the problem to determine the slope dependent current 
during the epitaxial growth process of a crystal surface.
This current is 
proportional to $\delta=p_+-p_-$, where $p_\pm$ are the probabilities
for an atom landing on a terrace to attach to the ascending ($p_+$)
or descending $(p_-)$ step. If the landing probability is spatially
uniform, the current is proved to be proportional to the average (signed)
distance traveled by an adatom before incorporation in the growing
surface. The phenomenon of slope selection is determined by the
vanishing of the asymmetry $\delta$. We apply our results to the case
of atoms feeling step edge barriers and downward funnelling,
or step edge barriers and steering.  In the general case,
it is not correct to consider the slope dependent current $j$ as a
sum of separate contributions due to different mechanisms.

\vskip 0.5\baselineskip

\selectlanguage{francais}
\noindent{\bf R\'esum\'e}
\vskip 0.5\baselineskip
\noindent
{\bf Courants de surface et s\'election de la pente en croissance
cristalline}
La croissance \'epitaxiale d'une surface cristalline peut \^etre
caract\'eris\'ee par un courant de surface $J$, dont
la partie $j$ qui d\'epend de la pente est \'etudi\'ee. Celle-ci
est proportionnelle \`a $\delta =p_+ - p_-$, o\`u $p_\pm$
sont les probabilit\'es qu'un atome d\'epos\'e sur une terrasse se colle
\`a la marche montante ($p_+$) ou descendante ($p_-$).
Si la probabilit\'e spatiale d'atterrissage est uniforme, le courant est
aussi proportionnel \`a la distance moyenne (avec signe) parcourue
par chaque atome. Le ph\'enom\`ene de la s\'election de la pente est
d\'etermin\'e par la condition $\delta=0$. Les r\'esultats ainsi obtenus
sont appliqu\'es aux cas barri\`ere de marche plus downward
funnelling et barri\`ere de marche plus braquage (steering). Dans le cas
g\'en\'eral, le courant $j$ ne peut pas \^etre consid\'er\'e comme la somme
de contributions s\'epar\'es dues aux diff\'erents m\'ecanismes.

 \vskip 0.5\baselineskip
\noindent{\small{\it Mots-cl\'es~:} Croissance cristalline~; 
Courant de surface~; Diffusion}}
\end{abstract}
\end{frontmatter}


\selectlanguage{english}

\section{Introduction}

Molecular Beam Epitaxy is a well known and widespread technique
to growth layers of metal and semiconductor crystals.
The growth process of a high symmetry crystal surface 
can be described through a ballistic flux of particles impinging
on the growing surface and the following thermal diffusion of adatoms,
which finally attach to a preexistent step or nucleate new 
terraces~\cite{PV_book,review,MK_book}.

One crucial aspect of the growth process is its possible unstable
character, due to deterministic mechanisms 
which prevent the growing surface to remain flat.
In this paper we focus on mound formation 
and we will discuss its description via a mesoscopic surface current
$J$, which enters in the evolution 
equation $\partial_t z = Ft -\partial_x J$, where $F$ is the
flux of incoming particles and $z$ is the local height.
Experiments show that dynamics can produce coarsening, with the
formation of mound facets of constant slope $m^*$~\cite{CS}.

In the theoretical seminal 
papers~\cite{JV_seminal,ElkVil} the cause for mound formation was identified 
in the existence of step edge barriers, i.e. an additional barrier
hindering the descent of steps. Since then, many 
efforts~\cite{review} have been devoted to the formalization
of this idea and to the derivation of quantitative predictions
for the evolution of mounds. These efforts have often combined
phenomenological approaches with attempts of a rigorous derivation
of $J$ starting from the microscopic dynamics of adatons.

The main piece of the current $J$ is the so called slope
dependent current $j(m)$, depending on the
local slope $m$ of the surface. $J$ also contains terms depending on
higher order derivates, but the rising of the instability
and the possible formation of mounds with a constant slope $m^*$
only depend on $j(m)$. Therefore, it is natural that
special attention has been devoted to its determination.
Analytical approaches have followed two mainstreams:
coarse-graining procedures~\cite{PolVil,Krug,WE} to pass from step-dynamics to
mesoscopic dynamics (see next Section) and the evaluation of $j(m)$
through the average (signed) distance~\cite{KPS,AF} walked by adatoms 
before being incorporated in the growing crystal (see Section~5).

A recent paper by Li and Evans~\cite{Evans} has renewed the interest
on the slope dependent current. Authors claimed that standard phenomenological
continuum theories are inappropriate to describe mound slope
selection. Afterwards some of their claims have been 
corrected~\cite{erratum}, but
their work has shown that $j(m)$ should be evaluated
with great care. For these reasons, in the following we reconsider
the problem, giving a formulation as general as possible
for the slope dependent current and for the evaluation of the
selected slope $m^*$, determined by the condition $j(m^*)=0$.
We prove that the current $j(m)$ is proportional
to the asymmetry $\delta(L)=p_+(L)-p_-(L)$ between the probabilities 
for an atom deposited on a terrace of size $L$ to be incorporated
into the upper ($p_+$) and lower ($p_-$) steps.
This proof does not refer to any specific microscopic mechanism.
An important by-product is that,
in general, $j(m)$ can not be considered as a sum of contributions due
to separate mechanisms. Finally, we also discuss the alternative
formulation for the current in terms of the averaged (signed) distance
walked by adatoms before incorporation.

\section{The current}
\label{sec_current}

In the case of conserved growth (no desorption, no overhangs), 
a useful concept to study the dynamics of the surface is
the mesoscopic current $J$, entering via the evolution
equation $\partial_t z = Ft -\partial_x J$ for the local height $z$.
It is worth stressing that
the dynamics of the surface, and therefore the current $J$, are determined
by the dynamics of steps: adatoms 
enter only through their attachment (and detachment) rate to steps.

A piece of surface of slope $m$ looks differently, according to
the value of the slope. For large $m$ (Fig.~1a) we have a sequence
of all uphill or downhill terraces of size $L\simeq 1/|m|$.
For small $m$ (Fig.~1b) we have a mix of different types of terraces of
size $L\simeq \ell_D$, where $\ell_D$ is the nucleation length~\cite{review}: 
in this case, the average slope is determined by uncompensated uphill
and downhill terraces. The two pieces of surface look differently because nucleation 
on terraces prevent them being larger than $\ell_D$.

\begin{figure}
\centerline{
\includegraphics[width=12cm,clip]{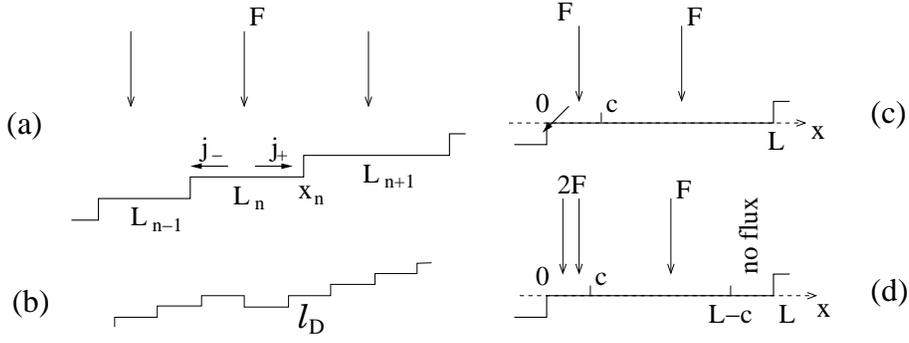}
}
\caption{(a,b) Schematic of a growing one dimensional surface
at (a) large and (b) small slope (figure b has been shrinked).
(c,d) Coordinates for the models studied in Section~\protect\ref{sec_dfs}:
(c) Downward Funnelling and Ehrlich-Schwoebel effect;
(d) steering and Ehrlich-Schwoebel effect.
}
\end{figure}

The analytical determination of the current $J$ starts from the
large slope case (see Fig.~1a).
In a one dimensional picture,
the flux $FL$ of atoms landing on the terrace splits 
in two currents $j_\pm(L)=FLp_\pm(L)$, where $p_\pm(L)$ are the
probabilities that an atom attaches to the ascending $(p_+)$ or
descending $(p_-)$ step. 
The velocity of the $n-$th step is simply equal to $[j_-(L_{n+1})+
j_+(L_n)]$. The sum $(p_+ + p_-)=1$, while
$\delta =(p_+ - p_-)$ defines the asymmetry $\delta$.

Following the method introduced in Ref.~\cite{PolVil}, the displacement
of step $n$ during the deposition of one monolayer 
can be approximated as
\be
x_n(t+1/F) - x_n(t) = -{1\over F} [j_-(L'')+j_+(L')] ,
\ee
where $L'=\um(L_n +L_{n-1})$ and $L''=\um(L_n +L_{n+1})$.
It is useful to sum the quantity $L_n$ to both sides, so as to have
\be
x_n(t+1/F) - x_n(t) +L_n = -\mbox{${1\over 2}$}
[L'' + L' -2L_n +L'\delta(L')-L''\delta(L'')] .
\ee
The quantity on the left
can be approximated as $(-1/mF)\partial_t z = (1/mF)\partial_x J$,
where $m=1/L$ is the slope.
The quantity on the right can be worked out using the
relations~\cite{PolVil,note_CG}
\bea
L'' + L' -2L_n &\approx &\mbox{${1\over 2}$}\left[ L^2\partial_{xx}L
+ L(\partial_x L)^2\right] \\
L'\delta(L')-L''\delta(L'') &\approx & (L'-L'')\partial_L[L\delta(L)] 
\approx  L\partial_x L \partial_L[L\delta(L)] . 
\eea

Finally, we get
\be
{1\over mF} \partial_x J = {1\over m} \partial_x \left[
-{L\partial_x L\over 4} + {L\delta(L)\over 2} \right] ,
\ee
so that the total current $J$ comes out to be
\be
J = \mbox{${1\over 2}$} FL\delta(L) - \mbox{${1\over 4}$}FL\partial_x L .
\label{eq_J}
\ee 

If we use the slope $m$, the second term takes the form
$\mbox{${1\over 4}$}F(\partial_x m)/m^3$, which was already found in
Refs.~\cite{PolVil,Krug}. 
In the following we will focus on the slope dependent part, 
\be
j = \mbox{${1\over 2}$} FL\delta(L) .
\label{eq_j}
\ee
We stress that Eq.~(\ref{eq_j}) is the most general form, not depending on any
assumption on the miscroscopic processes occurring at the surface.
It is valid for large slope, $m>1/\ell_D$ (Fig.~1a).
In the opposite limit,
$m<1/\ell_D$ (Fig.~1b), $j(m)$ is linear~\cite{PolVil} in $m$,
as also expected from symmetry reasons for a slope dependent current:
$j(m)=j(1/\ell_D)\ell_D m$, where $j(1/\ell_D)=\um F\ell_D \delta(\ell_D)$ 
is evaluated according to (\ref{eq_j}).

The condition of instability of the flat surface reads
$j'(0)>0$, i.e. $j(1/\ell_D)>0$, as trivially shown by a
linear stability analysis~\cite{note_stability}.
Finally, if mounds develop facets with constant slope $m^*$, the current 
must vanish on it: $j(m^*)=0$.

\section{Downward funnelling, step edge barriers and steering}
\label{sec_dfs}

Let us now consider the following model (see Fig.~1c). 
Atoms deposited within a distance $c$
from the descending step are incorporated in it
(downward funnelling, DF), while atoms deposited in the remaining
$(L-c)$ portion of the terrace diffuse freely, feeling an
additional (Ehrlich-Schwoebel, ES) barrier at the descending step.
Atoms deposited in the $c$ region give a trivial contribution
$Fc$ to $j_-$. As for the others,
we must solve the diffusion equation $F+D\rho''(x)=0$ for
$c<x<L$ and $\rho''(x)=0$ for $0<x<c$, with boundary conditions
$\rho'(0)=\rho(0)/\les$, $\rho(L)=0$, and $\rho(x),\rho'(x)$ continuous
in $x=c$. The quantity $\les\ge 0$ is the well known Ehrlich-Schwoebel
length and measures the additional barrier felt by an adatom 
in the sticking process to the descending step.
It is straightforward to get
\be
\rho'(0) = {F\over 2D}{(L-c)^2\over L+\les} , ~~~~~~~ 
\rho'(L) = -{F\over 2D}{(L-c)(L+c+2\les)\over L+\les} .
\ee

The contributions of atoms deposited in the $(L-c)$ region to the currents
$j_\pm$ are $D|\rho'(L)|$ and $D\rho'(0)$, respectively.
Therefore
\be
j_- = Fc + {F\over 2}{(L-c)^2\over L+\les}\equiv FLp_-(L) , ~~~~
j_+ = {F\over 2}{(L-c)(L+c+2\les)\over L+\les}\equiv FLp_+(L) .
\ee

Their sum gives the total flux of particles arriving 
on the terrace: $j_++j_-=FL$, while their (semi-)difference gives
the current $j=\mbox{${1\over 2}$}FL\delta(L)=\mbox{${1\over 2}$}
(j_+ - j_-)$, i.e.
\be
j ={F\over 2}{\les(L-2c)-c^2\over L+\les}
= {F\over 2}{\les(1-2mc) -mc^2\over 1+m\les} .
\label{J}
\ee
If $c=0$, we find the well known result $j=\mbox{${1\over 2}$}
FL\les/(L+\les)=\mbox{${1\over 2}$}F\les/(1+m\les)$.

Eq.~(\ref{J}) shows that it is not generally possible to write $j$
as a sum of two separate contributions, $j=j_{ES}+j_{DF}$, due to
the Ehrlich-Schwoebel (ES) effect and to the downward funnelling (DF),
respectively. In fact, it is instructive to consider the two 
limits $\les=0$ and $\les=\infty$.
For $\les=\infty$, $j_-=Fc$ and $j_+=F(L-c)$, i.e. $j_-$ is entirely
due to downward funnelling and $j_+$ to the ES effect. It seems natural
to write $j=j_{DF}+j_{ES}$, with $j_{DF}=-\um Fc$ and
$j_{ES}=\um F(L-c)$. For $\les=0$ we are induced to assume
$j=j_{DF}=-Fc^2/2L$. Therefore, if we want to write $j=j_{DF}+j_{ES}$,
the expression for $j_{DF}$ depends on $\les$: it is constant (not
depending on $m$) for large barriers and it is linear in $m$ for
weak barriers. The conclusion is that $j(L)$ should be handled as a whole.

Let us now consider a situation where the flux of particles on the
terrace is not uniform, see Fig.~1d. 
Because of steering effects~\cite{steering}, we assume that
all atoms destined to land within  distance $c$ from the ascending
step are steered and finally land in the $c$ region close to the
descending step (of the upper terrace). So, this region undergoes
an effective flux $2F$. Working out calculations~\cite{note_RW} 
similar to those made here above, we get the result
\be
j = {F\over 2}{\les L -2c(L-c)\over L+\les}
={F\over 2}{2c^2m + \les -2c\over 1+m\les} .
\ee

This current is always positive if $\les >2c$, or it changes sign
in $m^*=(1-\les/2c)/c$, if $\les <2c$ (see Fig.~2b, dashed line). 
However, in the latter case
$j'(m^*)>0$, which implies instability of the slope $m^*$~\cite{note_J}.
So, in this model there is no slope selection: either the surface is
always unstable or it is metastable~\cite{Amar}.

\section{Slope selection} 

First, let us apply the condition $j(m)=0$ to a model studied
by Li and Evans~\cite{Evans}:
\be
j_+ = F(L-c)P_+ , ~~~~
j_- = Fc + F(L-c)P_- . 
\label{Jevans}
\ee

\begin{figure}
\centerline{
\includegraphics[width=12cm,clip]{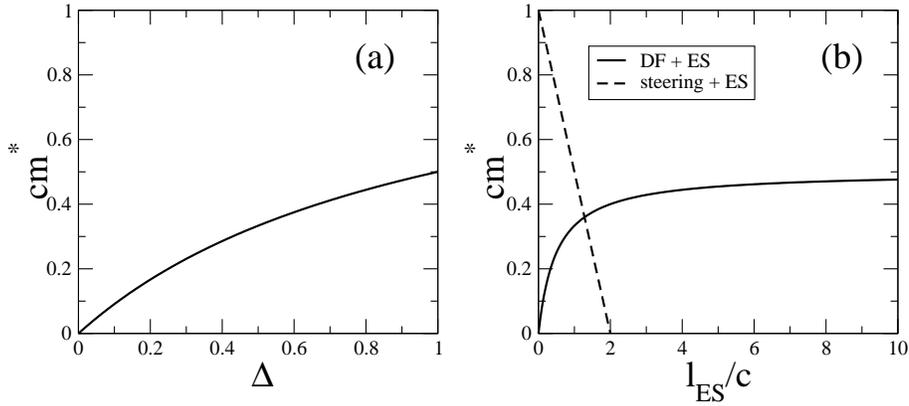}
}
\caption{(a) The selected slope $cm^*$ as a function of
$\Delta$, Eqs.~(\ref{Jevans}-\ref{ss_evans}).
(b) Full line: the selected slope $cm^*$ as a function of $\les/c$
(DF+ES model, Eq.~(\ref{eq_mstar})).
Dashed line: the slope $m^*$ such that $j(m^*)=0$ for the
steering+ES model (Section~\protect\ref{sec_dfs}). There is no
slope selection in this case, because $j'(m^*)>0$. }
\end{figure}

If $P_\pm$ are the probabilities for atoms deposited in the $(L-c)$ region 
to attach to the step, this model is similar to what we studied in 
the previous Section. For constant $P_\pm$, the condition 
$j_+=j_-$, i.e. $\delta=0$, 
gives the following expression for the selected slope $m^*$,
\be
m^* ={\Delta\over c(1+\Delta)} ,
\label{ss_evans}
\ee
where $\Delta=P_+-P_-$. Note that $\Delta$ is not the total asymmetry
$\delta$,
which vanishes for $m=m^*$, but the asymmetry for atoms deposited
in the $(L-c)$ portion of the terrace only.
In Fig.~2a we report $cm^*(\Delta)$. Eq.~(\ref{ss_evans})
perfectly matches the numerical results given in \cite{Evans}.

Let us now turn back to Eq.~(\ref{J}). In this case, we get
\be
m^* = {\les\over c\,(c+2\les)} .
\label{eq_mstar}
\ee
In Fig.~2b (full line) we plot $cm^*$ as a function of
$\les/c$. For small and large $\les/c$, $m^*=\ell_{ES}/c^2$ 
and $m^*=1/(2c)$, respectively. Therefore, for large ES barrier
the selected slope corresponds to a terrace size equal to $2c$,
as expected by a trivial compensation of DF and ES effects.

The use of (\ref{J}), as well as of (\ref{Jevans}), 
to find $m^*$ is limited by the constraint $m^*>1/\ell_D$.
For Eq.~(\ref{J}) this means $\les>c^2/(\ell_D-2c)\approx
c^2/\ell_D$; in the opposite limit, $\les< c^2/\ell_D$, 
$j(1/\ell_D)<0$ and no selected slope exists: 
this happens because the DF effect is so strong to
induce a stabilizing (downhill) current at all slopes.

\section{Another expression for the current}

The expression $j(L)=\um FL\delta(L)$ is the most general one
for the slope dependent current on a region of (large) slope $m=1/L$.
The quantity $\delta(L)=(p_+-p_-)$ measures the asymmetry
between the sticking probabilities to the upper and lower steps.
The flux $F$ is usually assumed to be spatially uniform, apart
from fluctuations.
However, as anticipated in the previous Section, 
because of atom-substrate interactions, steering 
effects may occur and
atoms are no more deposited uniformly on the terrace.
In spite of this, the espression  $j(L)=\um FL\delta(L)$ still
continues to be correct. In the Introduction we mentioned a
possible different evaluation of $j(m)$, through 
the average signed distance $d$ 
walked by an adatom, from the deposition to the incorporation site.
In the following we are proving the equivalence of the two expressions, if the
flux is uniform. If it is not uniform, the average distance $d$ is not
an appropriate quantity to determine $j$.

The convention is to take $d$ positive if the atom attaches to the
ascending step (see Fig.~1):
\be
d = {1\over L}\int_0^L dx [(L-x)\tilde p_+(x) -x \tilde p_-(x)]
= \int_0^L dx \:\tilde p_+(x) -{1\over L} \int_0^L dx
(\tilde p_+(x)+\tilde p_-(x))x ,
\ee
with $\tilde p_\pm(x)$ being the probabilities that an adatom deposited in $x$
attaches in $x=0$ ($\tilde p_-(x)$) and in $x=L$ ($\tilde p_+(x)$). 
Since $p_\pm=(1/L)\int_0^L dx \tilde p_\pm(x)$ and 
$(\tilde p_+(x)+\tilde p_-(x))=1$, we get
$d = Lp_+ -\um L = \um L\delta(L)$, so that
\be
j(L)=\um FL\delta(L) = Fd .
\label{j_d}
\ee

In a Kinetic Monte Carlo simulation, we can easily implement the
above expression and write
\be
j_{\mbox{\tiny KMC}} = F {N_r - N_l\over N_a}
\label{j_KMC}
\ee
where $N_{r,l}$ are the total hops in the uphill and downhill direction,
and $N_a$ is the total number of deposited atoms.
Formula (\ref{j_KMC}) was firstly introduced in Ref.~\cite{KPS} and 
has the merit to be valid for a surface of any slope.
At author's knowledge, a rigorous derivation and a comparison with
the mesoscopic current were missing.

It is worth stressing that Eqs.~(\ref{j_d}-\ref{j_KMC}) are no more
applicable if the landing probability is not spatially uniform.
In this case, $j(L)$ is no more
proportional to $d$, as shown by a trivial example: no downward funnelling
($c=0$) and infinite ES barrier, $\les=\infty$. The resulting current 
$j=\um FL$ can be written as $j=Fd$ if the flux is uniform: in that case,
$d=\um L$. Different spatial distributions of the landing atoms
modify $d$ and therefore the expression $j=Fd$, but do not
modify the correct expression $j=\um FL\delta$,
because the step dynamics would be unchaged.

\section{Final remarks}

In this paper we mainly focused on the slope dependent current
and the slope selection mechanism. We have shown that a correct
derivation of this current is possible and its general expression
has been found. The condition $j(m^*)=0$ determines the selected slope.
More generally, the shape $z_s(x)$ (or $m_s(x)=z'_s(x)$) 
of stationary solutions
depends on the vanishing of the full current 
$J$ in all the points, $J(m_s(x))\equiv 0$.
It is worthnoting that the conditions $j(m^*)=0$ and $J(m_s(x))\equiv 0$, 
when applied to a discrete or discrete-continuous
model, are valid only averaging $J$ on time scales not smaller than 
$1/F$~\cite{Evans}.

With regard to additional terms in the current, our derivation
in Section~\ref{sec_current} shows that a symmetry breaking term, having the form
$J=\partial_x A(m^2)$, appears naturally
when we coarsen step dynamics. This term
is the only term surviving in a plain
vicinal surface without any additional microscopic mechanism
(no step edge barriers, no nucleation, no thermal detachment, no
downward funnelling):
so, in this sense, it should be considered the most fundamental one.
Finally, $J$ should also contain at least 
a Mullins-like term, $J\sim\partial_{xx}m$, which
may be due to several mechanisms~\cite{Nato}.

However, even if the rigour in the derivation of the mesoscopic current 
has improved in the course of time, the evolution equation
$\partial_t z = Ft -\nabla\cdot \jb$ is not much more than a phenomenological
equation,
specially in two dimensions where additional problems linked
to step edge diffusion exist. 
The dynamics of a truly vicinal surface, which can be studied with much more
rigour~\cite{Grenoble}, shows that the full nonlinear equations
governing the growth process of a real system are far more complicated.



\section*{Acknowledgements}
Lively discussions with J.W. Evans are gratefully acknowledged.


\begin{thebibliography}{00}
\bibitem{note_CG}
The two expressions on the left are expanded at the lowest orders.
\bibitem{note_stability}
The equation $\partial_t z = -j(1/\ell_D)\ell_D\partial_{xx}z$
is solved by $z(x,t)=\exp(iqx+\omega t)$ with
$\omega(q)=j(1/\ell_D)\ell_D q^2$.
\bibitem{note_RW}
The use of a continuum diffusion equation at atomic scales
may be questionable. However, if a very precise numerical output
is not required, its use for evaluating the asymmetry $\delta$
is appropriate.
\bibitem{note_J}
If $J(m^*)=0$, the slope $m^*$ is stable (unstable) if
$J'(m^*)$ is negative (positive).
\end{thebibliography}
\end{document}